\documentclass[11pt,fullpage]{article}
\usepackage{authblk}
\usepackage{graphicx}

\begin{document}

\title{Universal Digital Services Through Basic Broadband}

\author[1]{Micah Beck\thanks{\tt mbeck@utk.edu}}
\author[2]{Terry Moore\thanks{\tt tmoore@icl.utk.edu}}
\affil[1]{Dept. of Electrical Engineering and Computer Science, University of Tennessee}
\affil[2]{Innovative Computing Laboratory, University of Tennessee}
\date{July 2021}                     %% if you don't need date to appear
\setcounter{Maxaffil}{0}
\renewcommand\Affilfont{\itshape\small}

\maketitle
\begin{abstract}
In addressing the universal deployment of digital services it is necessary to decide on the breadth of "universality" and the type of functionality of "digital services".
These two decisions are linked: certain desirable functionality does not lend itself to being implemented in some environments or with certain constraints.
In this paper we define universality as achieving a level of universality in digital service that can bridge the digital divide not only throughout highly industrialized societies but also across the globe and in the face of disruption.
We then argue that some of the characteristics of current Internet broadband service, in particular support for synchronous telepresence such as videoconferencing, is a barrier to implementation strategies that could enable cheap and resilient universal deployment of a more ``basic'' service.
\end{abstract}
\section{Universal Digital Services}
From its nineteenth century origins in the reform of the United Kingdom's national postal system~\cite{Rawnsley1999}, the concept of \textit{universal service} has been associated with idea of providing affordable telecommunication that connects together a nation's individuals and communities in ways expected to produce various social and economic benefits. The advent of the telegraph later in the century created a form of \textit{digital} service that attained widespread deployment with international reach for the asynchronous exchange of symbolic messages.
The term ``universal service'' was actually coined at the beginning of the twentieth century to describe the ambition of the analog telephony industry to make its particular form of telepresence, which mimics synchronous human interaction, affordable to everyone. Today, synchronous telepresence (audio and video) is implemented digitally, and the infrastructure that supports both synchronous and asynchronous digital service has converged into a global datagram delivery service---the Internet.

%Thus, it is a longstanding and widely held belief that the availability of digital services spanning the wide area has a positive impact on the social and economic circumstances of a community or a nation~\cite{barnett2014pentagon}.
But it has also long been evident that the concept of ``universal service''  encapsulates a kind of dilemma: other things equal, to make a service more universal beyond a certain point, the functionality it offers needs to be reduced, and conversely; thus, the potential size of the population served and the functional strength of the service tend to vary inversely. This is the \textit{universal service dilemma} (USD) that the current push for ``universal broadband'' must confront. 

In the core of the rich industrialized world, as well other areas of the world with substantial investment in infrastructure, digital services of all kinds are delivered through high bandwidth, low latency broadband Internet service. Current calls for ``universal broadband'' seek to make low cost access to this type of broadband service, which we call ``premium'' because it is necessary to support video telepresence, ubiquitous. Given the strong infrastructure assumptions that premium broadband service makes, the USD makes us doubt that this goal can be achieved.

The thesis of this paper is that {\em critical digital services can be cheaply and ubiquitously delivered in a less synchronous mode that makes weaker assumptions}.
The key to enabling this result is to deploy a ``basic'' broadband service that supports high bandwidth services but does not necessarily support telepresence applications.
We will present an argument that choosing to deploy basic broadband service universally is not widely understood because of an accident of technical history which has resulted in the conflation of basic and premium digital services, making it difficult to consider them as alternatives.

\section{End-to-End Signalling}
Devices in the wide area can interact in many different ways to create a variety of services.
They can store data, apply computation to it and transfer it to other devices that are adjacent in the network.
Like all Modern computer networks, the Internet follows a particular pattern: endpoints such as computers, tablets and mobile phones perform fully general functions, but the nodes in the interior of the network, including switches and routers, are mainly restricted to forwarding packets. One important recent deviation from this end-to-end architecture has been the deployment of security firewalls.

Factoring distributed service into a specialized communication component implemented by the network and a general component implemented at endpoints is referred to as an ``end-to-end'' service architecture~\cite{Saltzer:1984:EAS:357401.357402}.
Adherence to an end-to-end service architecture is widely credited with the Internet's vaunted ability to be deployed widely and to grow exponentially.
The capacity of wide area links and the capabilities of edge devices has also increased exponentially, leading to an environment in which many applications seek to leverage high bandwidth point-to-point data transfer.

An end-to-end service architecture requires that every pattern of interaction between network endpoints be expressed in terms of synchronous communication because storage and computation are ruled out within the network.
High bandwidth point-to-point data transfer in an end-to-end system architecture requires {\em end-to-end signalling}, which in the Internet is primarily implemented through the TCP protocol.
High bandwidth TCP data transfer requires both {\em very low network traversal time}  (or `` latency'') in end-to-end signaling and low error rates.
For this reason, supporting high bandwidth point-to-point data transfer using current broadband Internet connectivity is defined in terms of support for sufficiently low end-to-end latencies and error rates.

\section{Defining Basic Broadband}
As described above, the Internet architecture makes no distinction between datagram delivery for end-to-end TCP signaling and datagram delivery for actual content transfer. Consequently, it effectively conflates low latency end-to-end signalling and high performance point-to-point data transfer. This conflation is especially problematic if, in the face of the USD, our goal is to deliver the strongest possible service that can be made truly universal, i.e., made available at reasonable rates to \textbf{all} consumers/citizens, including those in low income, rural, remote, and high cost areas. 

Datagram delivery that is \textit{both} high bandwidth and low end-to-end latency---\textit{premium broadband}---is necessary \textit{only} for demanding types of \textit{synchronous telepresence}, such as voice calls, videoconferencing, and the sharing of multisensory virtual worlds. At the same time, the experience of wealthy, industrialized areas (i.e., areas that \textit{already} have a solid infrastructure base) shows that the level of investment needed to support premium broadband for a large and highly distributed user community is relatively high. Moreover, as we move away from this industrialized core, these costs tend to rise dramatically. Thus, if we take ``premium broadband for everyone'' as our goal, the USD suggests that we will have to dramatically narrow our definition of who counts as part of ``everyone'' in order to make claims of success plausible. 
% Of course, such a rhetorical approach to ``universal service'' is not unprecedented.

But once we recognize that high bandwidth delivery and low end-to-end latency (required for synchronous telepresence) are distinct and separable features of wide area digital services infrastructure, we can consider defining a form of digital service---{\em basic broadband}---that delivers the former but not necessarily the latter.
However, this does not tell us whether such a basic broadband service, with {\em weaker expectations} for quality of service but much broader availability,
%(since low latency Internet service does not actually provide {\em guarantees}) 
would be worth implementing.
This possibility raises two questions: 1) Could such a weaker service deliver value to users that is comparable to broadband as it is currently defined? and 2) Would it be substantially cheaper and easier to deploy, leading to levels of penetration approaching the truly universal?
Our answer in both cases is "yes".

%Now, just as synchronous wide area connections can be used to implement asynchronous communication ---a fact made obvious by the historical development of telegraphy--- the advent of dedicated (i.e. low latency and low error rate) Internet connections showed that the inverse was also true: high bandwidth datagram delivery can implement synchronous communication.

%This duality has given rise to the notion that the Internet represents not only the convergence of asynchronous delivery of large media objects (such as software and high quality multimedia recordings) and high bandwidth services (such as streaming delivery to such content), but also access to increasingly demanding types of synchronous telepresence.
%These range from text chat to voice calls and videoconferencing and sharing of multisensory virtual worlds.
%The recent sudden need to implement remote business and education due to the global pandemic has led to the call for ambitious levels of broadband service that can support such telepresence to be widely, or even universally, deployed in order to provide even the basic requirements of shared civic life.

\subsection{Data Logistics}
A digital service can be characterized by the observable behavior of end systems when certain stimuli are applied.
A behavioral characterization of this kind avoids specifying the internal state of end systems or the transfers of data over networks.
For instance, quality of service in the display of the frames of a video is independent of whether the video is stored at the endpoint or streamed over a network~\cite{Beck02anend-to-end}.

The ability to use data persistence, transfer and transformation (computation) in flexible ways to implement a certain functionality is the basis for a number of mechanisms that are commonly used to deliver digital services.
Examples are the use of caching to improve the access performance of distributed storage, of compression to improve the density of storage and of forward error correction to improve the accuracy of communication.
We refer to the application of this broad category of cross-resource optimizations as {\em data logistics}, in an analogy to the study of transportation  logistics.

The coupling of high bandwidth data transfer to low latency end-to-end signaling in broadband connectivity means that low latency datagram delivery can be used as a very powerful tool in implementing many valuable applications.
A familiar case in point is the delivery of digital content that is protected by copyright.
Solutions that rely on indefinite local storage of content use encryption which can be overcome through the application of processing.
Avoiding piracy requires intrusive software control over local storage and complex key management.
Reliance on streamed content enables the storage of data to be transient and it can be protected without intruding into the user-accessible storage resources of end systems.

Low latency end-to-end datagram delivery can be used to solve many other problems in very direct ways.
Installation of certain high performance software on heterogeneous end systems in a variety of edge environments can be challenging.
If software can be installed in a controlled data center and accessed over broadband connectivity, the installation problem can be finessed. 
Remote access to computation on demand then requires low latency communication between edge networks and the data center.

In these examples data logistics can sometimes be used to approximate or even replicate services whose current implementation relies on low latency datagram delivery.
The most important class of applications that cannot be approximated through logistical optimization can be broadly characterized as synchronous telepresence. 
The accurate recreation of remote sensory experiences requires low latency communication, although it may be possible to minimize the necessary bandwidth through compression and simulation.

Powerful techniques that are available for the application of data logistics to wide area digital services include caching, prestaging and replication of data, computation applied to data in transit and in situ as well as in dedicated edge servers~\cite{10.1145/1022594.1022596}.
Sophisticated learned models can be applied in edge devices to implement very low latency functionality, asynchronous store-and-forward communication, fault tolerance and load balancing applied to edge resources.
Some of these techniques require the application of sophisticated algorithms.
Others may require the adoption of standards throughout various parts of the edge-to-center continuum to enable interoperation between components implemented by different developers.
Data logistics, while powerful, is not always simple to apply.

\subsection{The Deployment Scalability Tradeoff}
Many necessary applications can be supported by basic broadband through data logistical optimizations.
However it is reasonable to ask whether an infrastructure that supports basic broadband can be implemented cheaply and easily enough to justify accepting weaker guarantees.

\begin{figure}[hbt] \centering
	\includegraphics[width=3in]{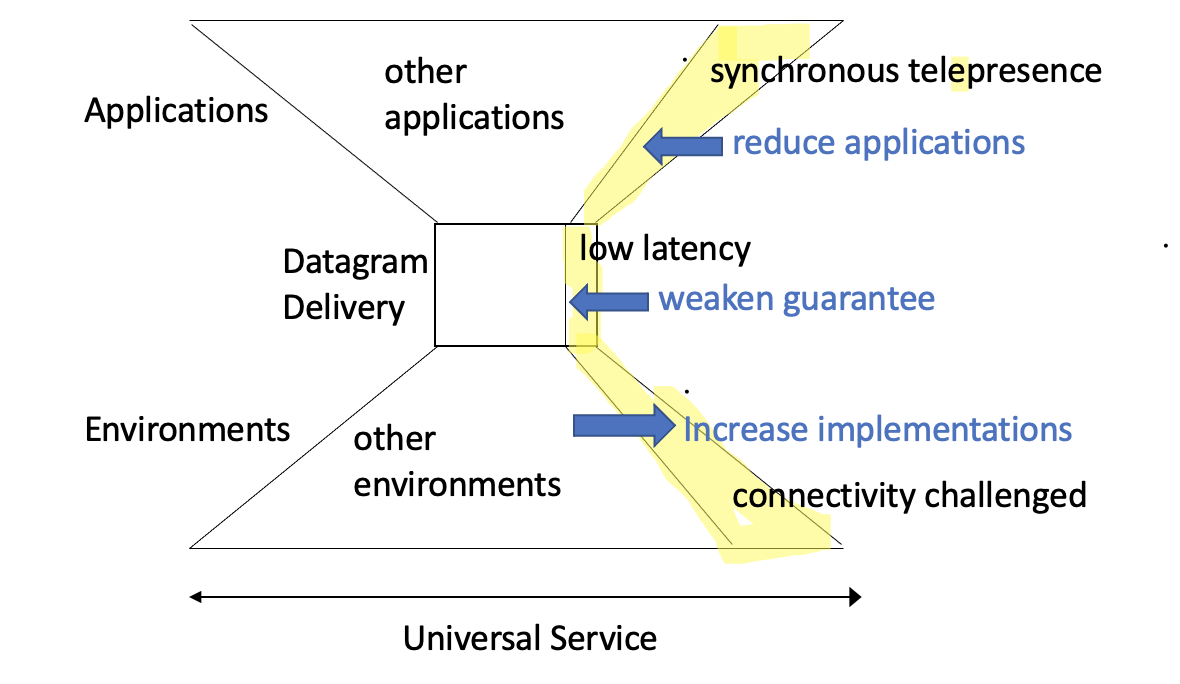}
	\caption{The Universal Service Hourglass illustrates how reducing supported applications through weakening of the common service can enable implementation in new environments. \label{hourglass}} 
\end{figure}

It is not possible to implement acceptable digital services without some bound on communication latency.
However if such bounds are very weak, then they enable a wide variety of implementation strategies.
The design process must take account of a principle known as the Deployment Scalability Tradeoff~\cite{beck2019hourglass}. It states that the ability of an infrastructure service to become ubiquitous correlates to the weakness of the guarantees it makes.
Following it, therefore, leads us to adopt the {\em weakest} service definition that is {\em still sufficient} to implement necessary applications, which in turn maximizes the possible implementations that can support those applications (as illustrated in Figure~\ref{hourglass}).

Basic broadband can be defined to have very {\em weak} bounds on the latency of communication.
While Voice over IP and telepresence require a latency bound on the order of 150ms~\cite{8653769}, many applications can tolerate average latency measured in hours.

A great variety of high bandwidth, high latency data transmission technologies and strategies can be used to implement basic broadband service.
One obvious choice is intermittent connectivity through satellites or other mobile relays including aerial drones.
Attempts have been made to use drones as relays to establish uninterrupted end-to-end communication without success.
Using intermittent connectivity significantly lowers the cost and engineering difficulty of such solutions.
More extreme is the use of data ferries, or vehicles equipped with data storage and local network connectivity, which can be used to move and relay stored data.
This approach is widely used to create ad-hoc, noninteroperable infrastructures, but it has not widely considered to be a type of network connectivity because it cannot support low latency end-to-end interactions.
However such techniques can support basic broadband services managed using data logistics~\cite{beck2019data}.

\section{A Basic Broadband Strategy}

There is ample evidence that basic broadband can be  successfully deployed widely, quickly and cheaply.
Experience with data logistics optimization has shown that many high value applications can be immediately implemented and that modern service management techniques can be used to increase the scope and variety of basic broadband applications.
Establishing standards would greatly speed and promote uniformity in the application environment.

Defining a common basic broadband service would in no way prevent the penetration of premium broadband service, which would automatically support basic connectivity.
The definition of basic broadband would however define a set of assumptions enabling universal delivery.
Critical applications such as e-government and public health and education could be implemented in a manner that is deliverable over basic broadband.
Emergency services such as disaster recovery could also be made more ubiquitous by requiring that they be supported by basic broadband. 

Part of the appeal of universal Internet connectivity is the idea that it creates a common service that can bind the entire world more closely together.
This notion may however contradict a fundamental fact of modern society and perhaps of physics: that end-to-end low latency datagram delivery requires too many strong assumptions to enable global universal deployment.
Remember that``universal'' telephone service was defined as affordable but minimal local service supporting only critical communication functions -- long distance was a premium service and international service even less available.
The Internet was never designed to to connect every shepherd on the Asian Steppe.
Trying to impose commonality where it does not naturally exist can have the unintended effect of marginalizing the very communities that we most want to include because they do not conform to a priori assumptions.

If end-to-end datagram delivery is too strong to be the basis for universal service, does that mean that all hope of global interoperability is lost?
Not if a more fundamental basis can be established for commonality in the implementation of distributed systems using data storage, transfer and computation.
If the low level mechanisms that are used to implement basic broadband applications are standardized then a variety of digital services can be deployed globally through adoption of those standards.
This suggests agreeing on a common virtual model of node resources (a node operating system) and local area connectivity to enable a more realistic basis for global deployment of services.
Such a strategy may not fit the business models of the corporations that currently seek to capture their customers within private Cloud-to-edge application ecosystems, but it may be a path to even greater global integration and interoperation.

\vspace{-.05in}
\section{Related Work}
The use of data logistics to deliver digital services in the wide area predates the Internet era.
File-based networks such as UUCP and BITNET used server-based store-and-forward and application routing and processing in transit to implement early services including email and bulletin boards.
Online service such as AOL and CompuServe also used server-based resources augmented with low-bandwidth, low latency telephone connectivity and high bandwidth, high bandwidth data transfer using magnetic or optical disks (data ferrying).
In each case there was a corresponding standard model of edge processing (Unix, IBM mainframe and IBM PC).

Data logistics were also brought into play to overcome the limitations of point-to-point communication using the client/server model of the early Internet.
FTP mirroring, Web caching, Content Delivery Networks and Cloud vendors all use data logistics to provide end users an architecture- and topology-independent view of distributed resources.
Over the past 20 years a number of stateful approaches to wide area networking have been developed to avoid the assumptions and constraints of end-to-end datagram delivery.
These have included Delay and Disruption Tolerant Networking~\cite{1204759} and  Content Centric Networking~\cite{8653769}.
Each of these has faced problems of implementation in an environment in which global Internet routing is assumed to be the universal basis of interoperability.

\vspace{-.05in}
\section{Current Broadband Initiatives}
As a matter of public policy, the goal of deploying broadband widely is to make necessary services available. 
We have argued that deployment of basic broadband service, without requiring synchronous telepresence to be supported, is sufficient to reach this goal.
Deploying premium broadband, and thus requiring support for synchronous telepresence, does more than what is minimally necessary.

The belief that the features of premium broadband service will enhance the lives of those it reaches is widespread.
A 2017 IEEE Internet Technology Policy Community White Paper 
% entitled "Options and Challenges in Providing Universal Access"
~\cite{Aledhari2018WhitePO}
cites the 17 UN Sustainable Development Goals~\cite{UNSDG} as motivation.
It claims that ``almost all of them require universal Internet access ... for their necessary development.''
Current initiatives to expand the penetration of broadband do not make the distinction between basic and premium service.
Low latency end-to-end datagram delivery is considered an implicit characteristic of broadband connectivity.

As an example, the California Digital Divide Innovation Challenge, announced in February 2021, offers US\$1M in prize money for "the boldest, most revolutionary proposals to eliminate the digital divide and expand high-speed internet access to all Californians"~\cite{CalInnChallenge}.
Candidate solutions must provide 
a cost of no more than US\$15/month per household and must implement 100 Mbps synchronous upload and download speed.
The call does not specify low latency end-to-end datagram delivery, but solutions providing basic broadband connectivity were deemed to be out of scope.

The "Accessible, Affordable Internet for All Act", includes a definition of the broadband connectivity which is more explicit in requiring support for telepresence:
"...broadband service with a download speed of at least 100 megabits per second, an upload speed of at least 100 megabits per second, and a latency that is sufficiently low to allow real-time, interactive applications."~\cite{HR7302}

The reasons and motivation for including support for telepresence in the definition of broadband connectivity are easy to understand and are appealing.
The distinction between basic and premium broadband has never been formalized, and many valuable applications are currently implemented using premium features.
The market-driven history of Internet application development leads many end users to fear that unless they share the highest available level of connectivity they will be excluded from future participation.
Some fear that once they have access to basic broadband the ambition to make premium broadband universal will be forgotten.
These objections can all be addressed through appropriate regulation.
Implicitly including premium features in the definition of broadband may unnecessarily impede universal deployment. 

Whenever we set out to achieve a goal through technological means we make choices.
Some of these choices are influenced by historical circumstances, others by market or regulatory forces, or by personal or professional biases.
In every case there are effects other than our original goal.
The Deployment Scalability Tradeoff~\cite{beck2019hourglass} tells us that one effect of attempting to achieve a goal that is stronger (more ambitious) than necessary, may be a loss of widespread adoption of the ultimate solution.
Our good intentions must be matched by a clear-eyed understanding of the fundamental trade-offs.
Adopting a requirement that has the effect of imposing constraints may hinder achievement of necessary goals.

\section{Conclusions}

In the current implementation of broadband digital services, high bandwidth delivery of content is linked to low latency end-to-end signalling.
This has resulted in the conflation of these two different dimensions of "quality of service" in the implementation of universal broadband service.
Many critical applications can be implemented without an assumption of low latency end-to-end datagram delivery through the use of ``data logistics".
Most popular services that {\em require} low latency end-to-end data delivery can be characterized as various forms of ``synchronous telepresence''.

{\em Basic broadband} is a form of broadband service which does not require support for synchronous telepresence.
Experience gives us reason to believe that basic broadband can be cheaply and easily deployed and can support many critical and valuable digital services.
This suggests that even large scale investment designed to increase the penetration of premium broadband may not be successful in closing the digital divide in access to digital services.
In contrast, universal deployment of basic broadband is within reach.

\bibliographystyle{ieeetr}
\bibliography{ebpbib,hglass,goodit,MASS,expeditions}

\end{document}